\documentclass[preprint,12pt]{elsarticle}
\usepackage{graphicx}
\usepackage{amssymb}
\usepackage{multirow}
\usepackage{makecell}
\usepackage{url}
\usepackage{xcolor}
\usepackage[normalem]{ulem}
\usepackage{lineno}
\usepackage{hyphenat}

\journal{International Journal of Heat and Mass Transfer}
\biboptions{sort&compress}

\begin{document}

\begin{frontmatter}

\title{Silicon emissivity as a function of temperature}

\author{Marcio Constancio Jr}
\address{Divis\~ao de Astrof\'isica, Instituto Nacional de Pesquisas Espaciais, S\~ao Jos\'e dos Campos, SP, 12227-010,Brasil}

\author{Rana X. Adhikari}
\address{LIGO Laboratory, California Institute of Technology, Pasadena, CA 91125, USA}

\author{Odylio D. Aguiar}
\address{Divis\~ao de Astrof\'isica, Instituto Nacional de Pesquisas Espaciais, S\~ao Jos\'e dos Campos, SP, 12227-010,Brasil}

\author{Koji Arai}
\address{LIGO Laboratory, California Institute of Technology, Pasadena, CA 91125, USA}

\author{Aaron Markowitz}
\address{LIGO Laboratory, California Institute of Technology, Pasadena, CA 91125, USA}

\author{Marcos A. Okada}
\address{Divis\~ao de Astrof\'isica, Instituto Nacional de Pesquisas Espaciais, S\~ao Jos\'e dos Campos, SP, 12227-010,Brasil}

\author{Chris C. Wipf}
\address{LIGO Laboratory, California Institute of Technology, Pasadena, CA 91125, USA}

\begin{abstract}

In this paper we present the temperature-dependent emissivity of a silicon sample, estimated from its cool-down curve in a constant low temperature environment ($\sim 82 K$). The emissivity value follow a linear dependency in the 120-260 K temperature range. This result is of great interest to the LIGO Voyager gravitational wave interferometer project since it would mean that no extra high thermal emissivity coating on the test masses would be required in order to cool them down to 123 K. The results presented here indicate that bulk silicon itself can have sufficient thermal emissivity in order to cool the 200 kg LIGO Voyager test masses only by radiation in a reasonable short amount of time (less than a week). However, it is still not clear if the natural emissivity of silicon will be sufficient to maintain the LIGO Voyager test masses at the desired temperature (123 K) while removing power absorbed by the test masses. With the present results, a black coating on the barrel surface of the test masses would be necessary if power in excess of 6 W is delivered. 
However, the agreement we found between the hemispherical emissivity obtained by a theory of semi-transparent Silicon and the obtained experimental results makes us believe that the LIGO Voyager test masses, because of their dimensions, will have effective emissivities around 0.7, which would be enough to remove about 8.6 W (7.5 W) for a shield at 60 K (80K). This hypothesis may be confirmed in the near future with new measurements.

\end{abstract}

\begin{keyword}
LIGO \sep Silicon \sep thermal emissivity \sep gravitational waves

\end{keyword}

\end{frontmatter}

\section{Introduction}
\label{Sec:intro}

Since the inauguration of gravitational wave astronomy in 2016, several detections have been announced by the LIGO-VIRGO Scientific Collaboration (LVC)\cite{GW150914,GW170817,GWTC-1}. From now on, a new way to observe the Universe is open, with expectations ranging from completely new discoveries to the multi-messenger astronomy with electromagnetic counterpart and even neutrinos. 

Currently, advanced LIGO\cite{Aasi_aLIGO2015} and advanced Virgo\cite{AdvancedVirgo}, which are the $2^{nd}$ generation of the LIGO and Virgo detectors, have regularly operated searching for these signals. LIGO-India\cite{MemoLIGO-India} and KAGRA\cite{Kagra} are expected to join  this journey within the next few years.

While the $2^{nd}$ generation detectors are in operation, new detectors have being planned. These include the Einstein Telescope (ET) \cite{EinsteinTelescope2010} , LIGO Voyager\cite{adhikari2020cryogenic} and Cosmic Explorer detectors \cite{WhitePaper2018}. Between the last two, LIGO Voyager brings great technological challenges since it is a cryogenic update in LIGO's current facilities.

In the Voyager version, Silicon-made mirror and suspensions will operate at cryogenic temperatures due to some excellent properties, such as low mechanical loss in bulk silicon\cite{McGuigan1978} and other optical properties at wavelengths of 1.5 - 2.5 $\mu$m \cite{Rowan2003,Adhikari2014,Keevers1995}. Also, Silicon has a zero crossing in thermal expansion coefficient around 123 K, which can suppress thermoelastic noise at this temperature \cite{Swenson1983,Middelmann_2015,Kim2018}. The thermal expansion coefficient goes from -0.339 at 100 K to 2.618 at 300 K \cite{Swenson1983}.

However, to keep both mirror and suspensions at this temperature, about 10 W of power deposited by the interferometer laser light must be extracted from the test mass. This can be done through radiation if the emitting surface has sufficiently high thermal emissivity. High emissivity coating such as DLC - Diamond-Like Carbon coating \cite{Sakakibara2014} and Acktar Black coating \cite{AcktarBlack2} have been proposed to improve the barrel's thermal emissivity. The emissivity of diamond like carbon was found to follow the linear fitting $\epsilon_{DLC} = 0.3 (T/300K)$, which has lower values than the ones obtained for our sample. And the emissivity of Acktar Black coating is $>$ 0.98 (in the 3-10 $\mu$m range) and $>$ 0.93 (in the 3-30 $\mu$m range). However, we need to know its emissivity in the 3-100 $\mu$m range or as a function of temperature to really be able to compare its performance with our experimental results. Surface oxide layers can also increase the emissivity properties. Relevant references are \cite{Cesarini,LiVoti2018,Larciprete,Hassan,LiVoti2015}. However, any additional coating will deliver extra amounts of thermal noise to the mirror. 

In this paper we measure the emissivity of a bulk Silicon sample and show that it may have a temperature-dependent thermal emissivity sufficiently high in order to avoid the use of any extra coating on the barrel.

\section{Experiment: Sample and Setup}
\label{Sec:Exp}

In order to measure the total temperature-dependent emissivity of Silicon ($\epsilon(T)$), a bulk Silicon sample was radiatively cooled down to 123 K while having its temperature monitored by very thin thermocouples thermometers (0.0799 mm in diameter).

\begin{figure}[!ht]
\centering
\includegraphics[width=0.65\linewidth]{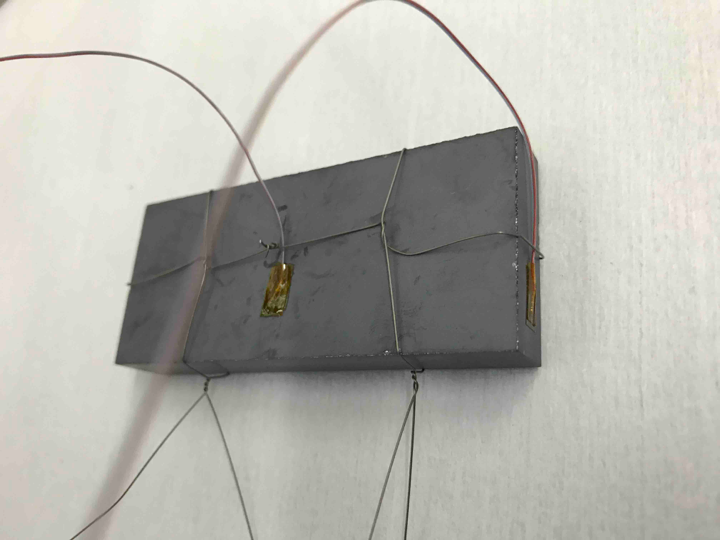}
\caption{Silicon sample used in this experiment. The sample has not a mirror finishing surface. T-type thermocouples and the Ti-6Al-4V wires are shown.}
\label{fig:Si-sample}
\end{figure}

The specimen used in this experiment is made of undoped, magnetic Czochralski\cite{Czochralski1918} grown silicon with a resistivity (as quoted by the manufacturer) of 4360 $\Omega$.cm. It is 70.00 mm $\times$ 30.00 mm $\times$ 10.35 mm rectangular parallelepiped shaped and weights 50.53 g. As shown in figure \ref{fig:Si-sample}, it has a rough finishing surface. The sample belongs to LIGO's Caltech group but the experiment was performed at Instituto Nacional de Pesquisas Espaciais (INPE),Sao Jose dos Campos,Sao Paulo state, Brazil.

\begin{figure}[!hb]
\centering
\includegraphics[width=0.7\linewidth]{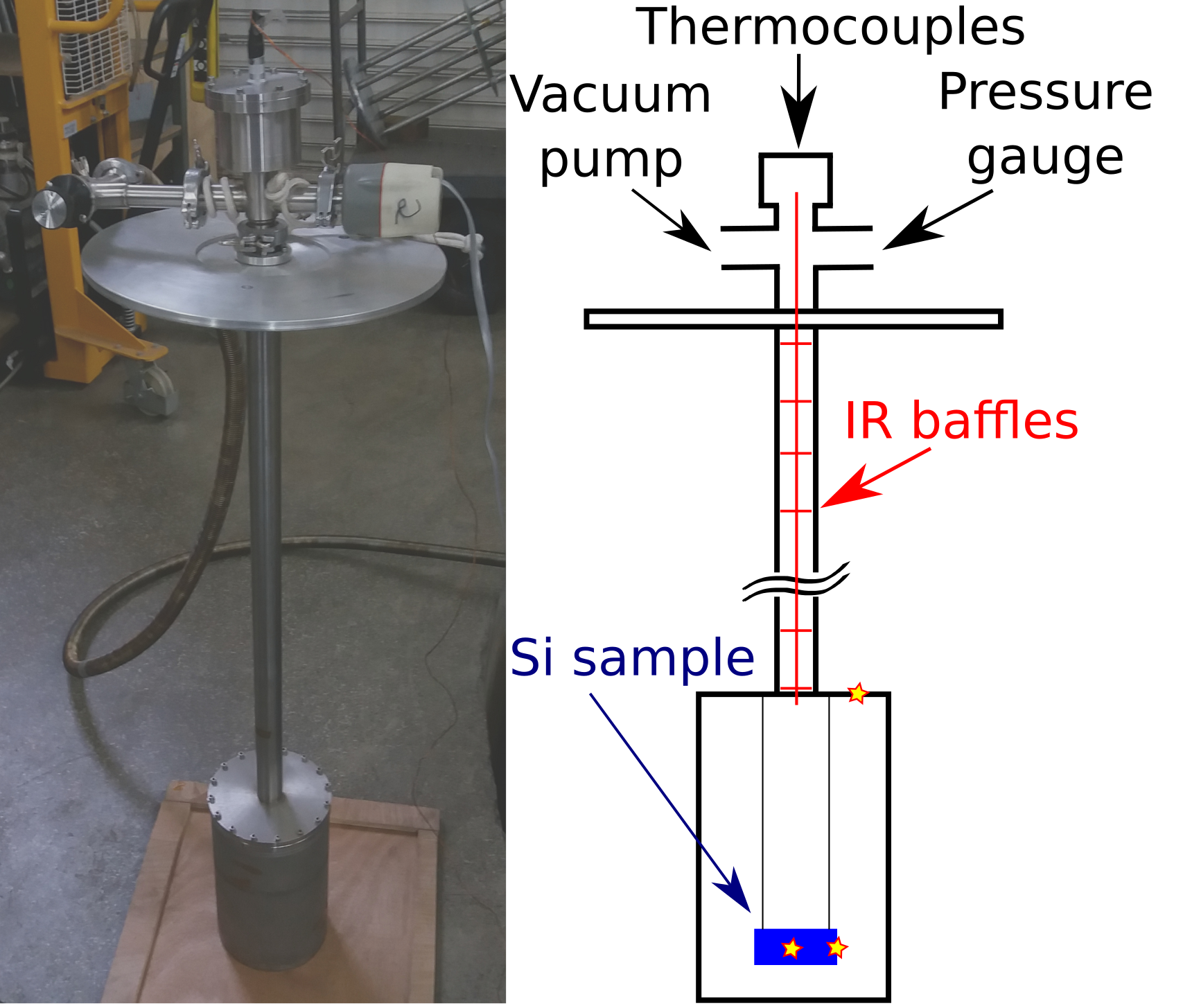}
\caption{Cryostat. A picture of the real one (left) and a diagram with details of the experiment (right).The stars show the position where the thermocouples were installed.}
\label{fig:Cryostat}
\end{figure}

The experiment was performed in a cryostat under high vacuum ($< 10^{-7}$ mbar). This means convection and air conduction could be neglected. The cryostat and a simplified diagram are shown in figure \ref{fig:Cryostat}.

The sample was hang from the top of the dewar by four 0.29 mm thick, 20 cm long Ti-6Al-4V wires. Although there was a large gradient along the wires, the heat transfer by them was 100-300 times lower than the dominant heat transfer by radiation in the 123-300 K temperature range. Three T-type thermocouples with 0.0799 mm in diameter (AWG40) were used to monitor the temperature, two of them were attached to the Si sample and the other in the top of the dewar (their position are shown with stars in figure \ref{fig:Cryostat}).

In order to create a sudden cryogenic environment, the dewar was rapidly immersed in a liquid nitrogen bath ($LN_2$). The amount of liquid was enough to cool the dewar down and to guarantee that it would be surrounded by $LN_2$ during the whole experiment.
Finally, nine infrared baffles were used to decrease the amount of heat coming from the top of the cryostat, which was kept at room temperature. In order to check for reproducibility, the experiment was performed twice.

\section{Results and discussions}
\label{Sec:result}
\subsection{Analysis of heat transfer processes involved}
Heat can be transferred from the Si sample to the cold dewar walls by the following processes: radiation, gas thermal conductivity, gas thermal convection, and metal thermal conductivity (by the suspension wires and thermocouple wires). The values of power at 300 K and 123 K for these various processes are, respectively, the following:
Radiation: 1.4 W (at 300K), 3.4$\times 10^{-2}$ W (at 123K); gas thermal conductivity: 4.1$\times 10^{-5}$ W (at 300K), 8.6$\times 10^{-6}$ W (at 123K); gas thermal convection: negligible; thermal conductivity of the suspension wires: 3.0$\times 10^{-3}$ W (at 300K), 3.3$\times 10^{-4}$ W (at 123K); thermal conductivity of the thermocouple wires: 1.8$\times 10^{-3}$ W (at 300K), 4.1$\times 10^{-4}$ W (at 123K).
Therefore, radiation is by far the dominant thermal transfer process. It is about 300 times and 46 times larger than all the other thermal transfer processes at 300 K and 123 K, respectively. So in the worse case (123K), all the other processes represent only about 2 percent.

\subsection{Experimental results}

\begin{figure}[!ht]
\centering
\includegraphics[width=1\linewidth]{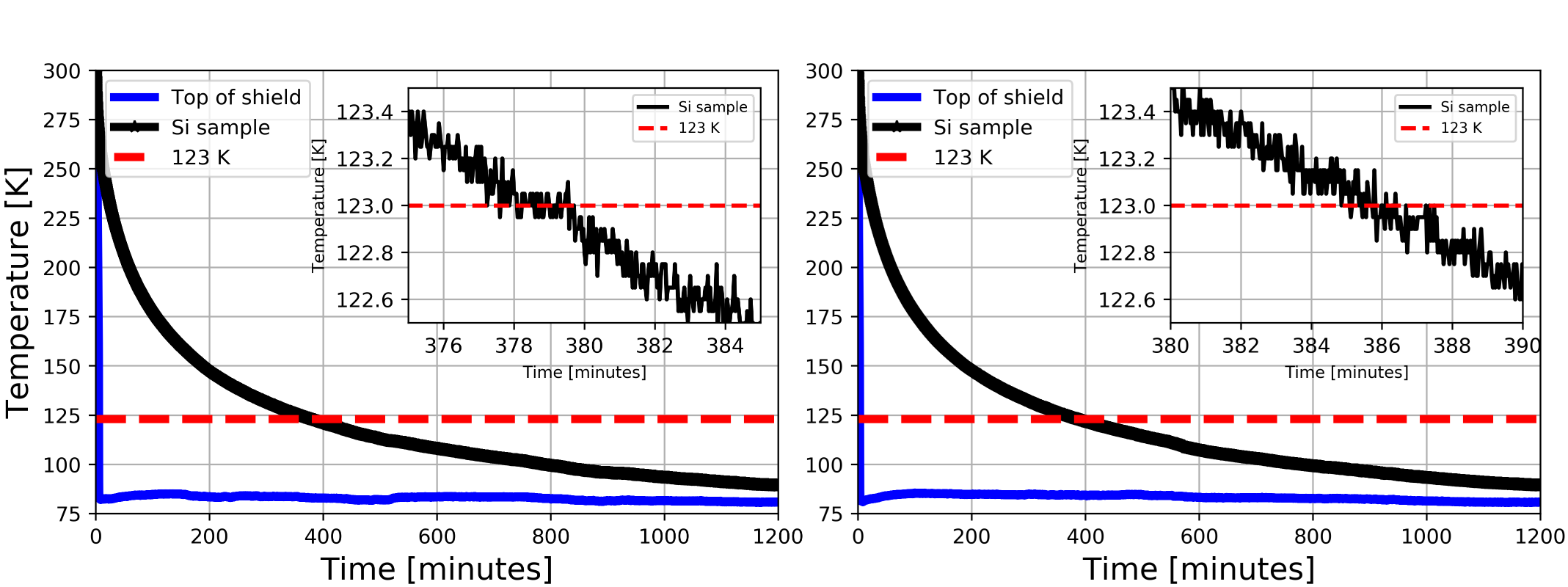}
\caption{Cooling down curve from run\# 1 (left) and run\# 2 (right). The inset shows the time when the Silicon sample crossed the target 123 K temperature and its level of noise.}
\label{fig:Coolingdown}
\end{figure}

As mentioned above, the experiment was performed twice in order to check for reproducibility. The top of the dewar achieved 82.5(84.2) K in about 6(4) minutes after the full immersion. This ensures that the whole dewar was cold at this time. The final cooling down curves for runs 1(2) are shown in figure \ref{fig:Coolingdown}. The inset shows the time when the sample crossed the target 123 K temperature and shows the ``noise'' created by a digital reading. Because the temperature reading is digital, the reading keeps going back and forth, instead of showing a monotonic behavior. 

Considering that all the energy leaving the specimen is purely radiative (convection, gas and wire conduction are negligible), the transient heat transfer (or energy balance) equation can be written as:

\begin{equation}
\label{eq:transient1}
m_{Si} C_{P}(T) \frac{dT}{dt} = A_{Si} \epsilon(T) \sigma (T^4 - T_{sh}^4)
\end{equation}

where $\epsilon(T)$ is the temperature-dependent total emissivity, $C_{P}(T)$ is the temperature-dependent heat capacity of Silicon\cite{Flubacher1959}, $\sigma$ is the Stefan-Boltzmann constant (= 5.6697$\times 10^{-8}~W.m^{-2}.K^{-4}$),$T_{Sh}$ is the temperature of the shield (dewar) and $m_{Si}$, $A_{Si}$ and T are, respectively, the mass, area and temperature of the specimen (which is supposed to be uniform, given the slow cooling rate and the high thermal conductivity of silicon). For a semi-transparent sample, which is the case, as discussed in section 3.3, the use of the sample surface area only makes sense for the purpose of defining an effective emissivity.

The emissivity can be calculated from the inverse problem by\cite{PENG2018}:

\begin{equation}
\label{eq:transient2}
\epsilon(T) = \frac{m_{Si}C_{P}(T)\frac{dT}{dt}}{A_{Si}  \sigma (T^4 - T_{sh}^4)}
\end{equation}

The derivative $\frac{dT}{dt}$ can be obtained from the data in figure \ref{fig:Coolingdown}, making the total temperature-dependent emissivity calculation feasible. However, since the data is noisy, because the temperature reading is digital (the reading keeps going back and forward), calculating the derivative from raw data can affect its accuracy. So, a few different approaches were used to understand the data. They are described in the following subsections.

\subsubsection{First data analysis approach: Savitzky-Golay (SG) filtering}
\label{subsub:SG}

\begin{figure}[!ht]
\centering
\includegraphics[width=1.0\linewidth]{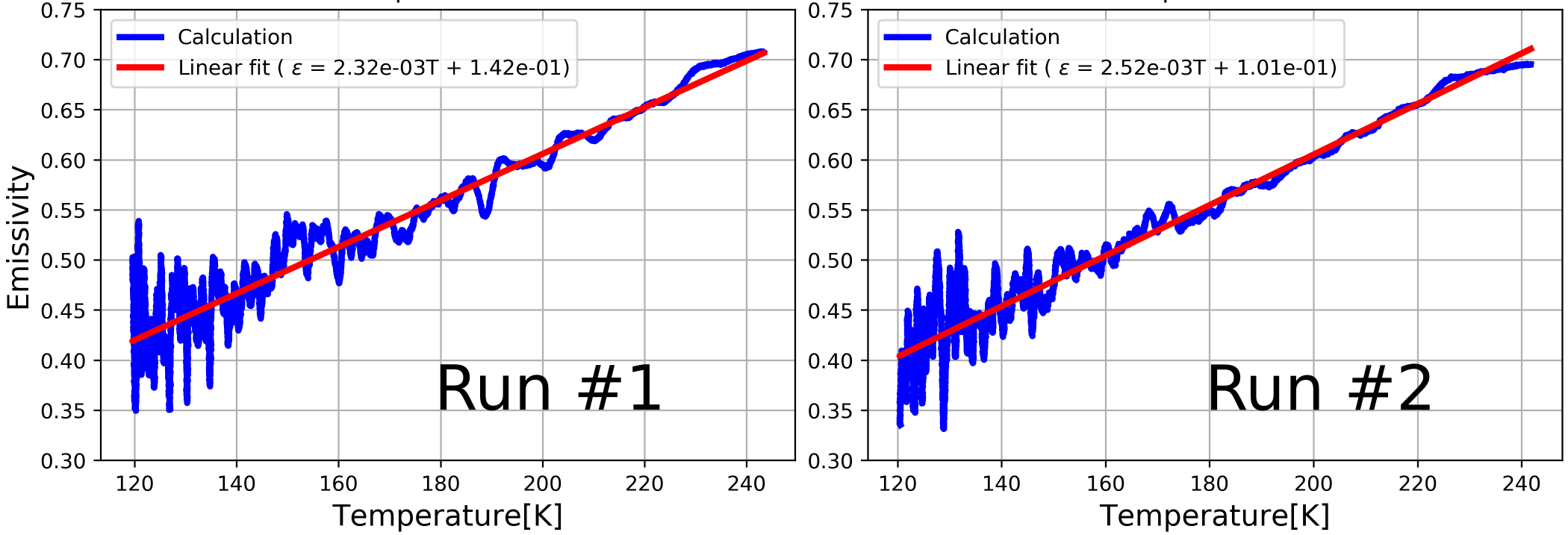}
\caption{Calculated emissivity for runs \#1 and \#2 by mean of SG filter. A mean value between these linear regression gives $\epsilon(T) = 2.42 \times 10^{-3}$ T $ + 1.22\times 10^{-1}$.}
\label{fig:Calc-emissivity_SG}
\end{figure}

In the first approach, we used a Savitzky-Golay (SG) filter, which is a digital filter based on a simplified least-squares fit convolution \cite{SG_1964}. SG uses a low-order polynomial in a smoothing window to fit the data. In this paper the window size is 201 and the polynomial degree is 2.
Scipy\cite{Scipy_savgol} package in Python has a SG-function which gives the derivative of smoothed data directly. The curve $\frac{dT}{dt}$ obtained from this method was used as input in equation \ref{eq:transient2}. The emissivity calculated from this method is shown in figure \ref{fig:Calc-emissivity_SG} for both runs. The red curve is a linear regression of the data and the mean between the both runs gives $\epsilon(T) = 2.42 \times 10^{-3}  T + 1.22 \times 10^{-1}$.

\subsubsection{Second approach: averaging the digital readings}
\label{subsub:smoo_der}

Another approach we used was a calculation where a simplified average value was calculated from a set of points of the original array. A new smoothed array $\bar{T}$ was created from the original data ($T$), which is denoted as ($T_{1},T_{2},T_{3},...T_{i})$. Every element $\bar{T_{j}}$ results from the average value of $T$ ranging from $i-n$ to $i+n$ of the original data, as follow:

\begin{equation}
\label{eq:Smooth}
\bar{T_{j}} = \frac{1}{(2n+1)}\sum^{k = (i + n)}_{k = (i - n)} T_{k} 
\end{equation}

From this analysis, one can expect that $\bar{T_{j}}$ has length equal to $len(T)-2n$. In this analysis, n = 150, which means that the first 301 points (i-n to i+n) from the original data were used to calculate T(151), which is the first point of this array or $\bar{T_{1}}$ (T(2) to T(302) will produce $\bar{T_{2}}$, and so on). The ``derivative'' was calculated from this smoothed data by the difference $\frac{\bar{T}_{(i-n)}-\bar{T}_{(i+n)}}{t_{(i+n)}-t_{(i-n)}}$. As before, the calculated $\frac{dT}{dt}$ was used as input in equation \ref{eq:transient2}.

\begin{figure}[!ht]
\centering
\includegraphics[width=1.0\linewidth]{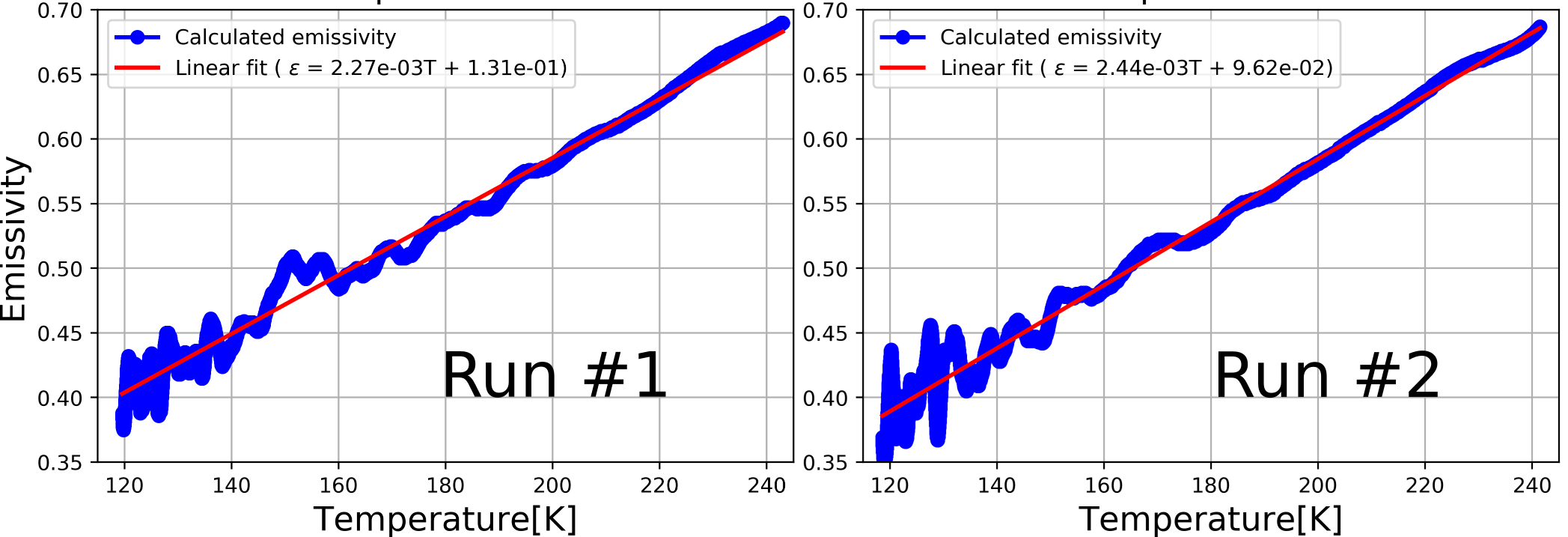}
\caption{Calculated emissivity for runs \#1 and \#2 by mean of a simplified average smooth. A mean value between the linear regression of both runs gives $\epsilon(T) = 2.36 \times 10^{-3}$ T $ + 1.14\times 10^{-1}$.}
\label{fig:Calc-emissivity_simp_aver}
\end{figure}

The emissivity calculated through this method is shown in figure \ref{fig:Calc-emissivity_simp_aver} for both runs. The red curve is a linear regression of the data and the mean value between the both runs gives $\epsilon_{Si}(T)$ = 2.36 $\times 10^{-3} $ T + 1.14 $\times 10^{-1}$.

\subsubsection{Third approach: Curve fitting}
\label{subsub:fitting}

Rather than smoothing (by averaging), as described above, another approach was tried to better understand the data and the results. Raw data was fitted to theoretical curves before any calculation was performed. Two different functions were used, a second order exponential function (${T(t) = a.e^{-b.t} + c.e^{-d.t} + e}$) and a $10^{th}$ order polynomial function ($T(t) = a.t^{10} + b.t^{9} + c.t^{8} + d.t^{7} + e.t^{6} + f.t^{5} + g.t^{4} + h.t^{3} + i.t^{2} + j.t + k$).

\begin{figure}[h]
\centering
\includegraphics[width=1.0\linewidth]{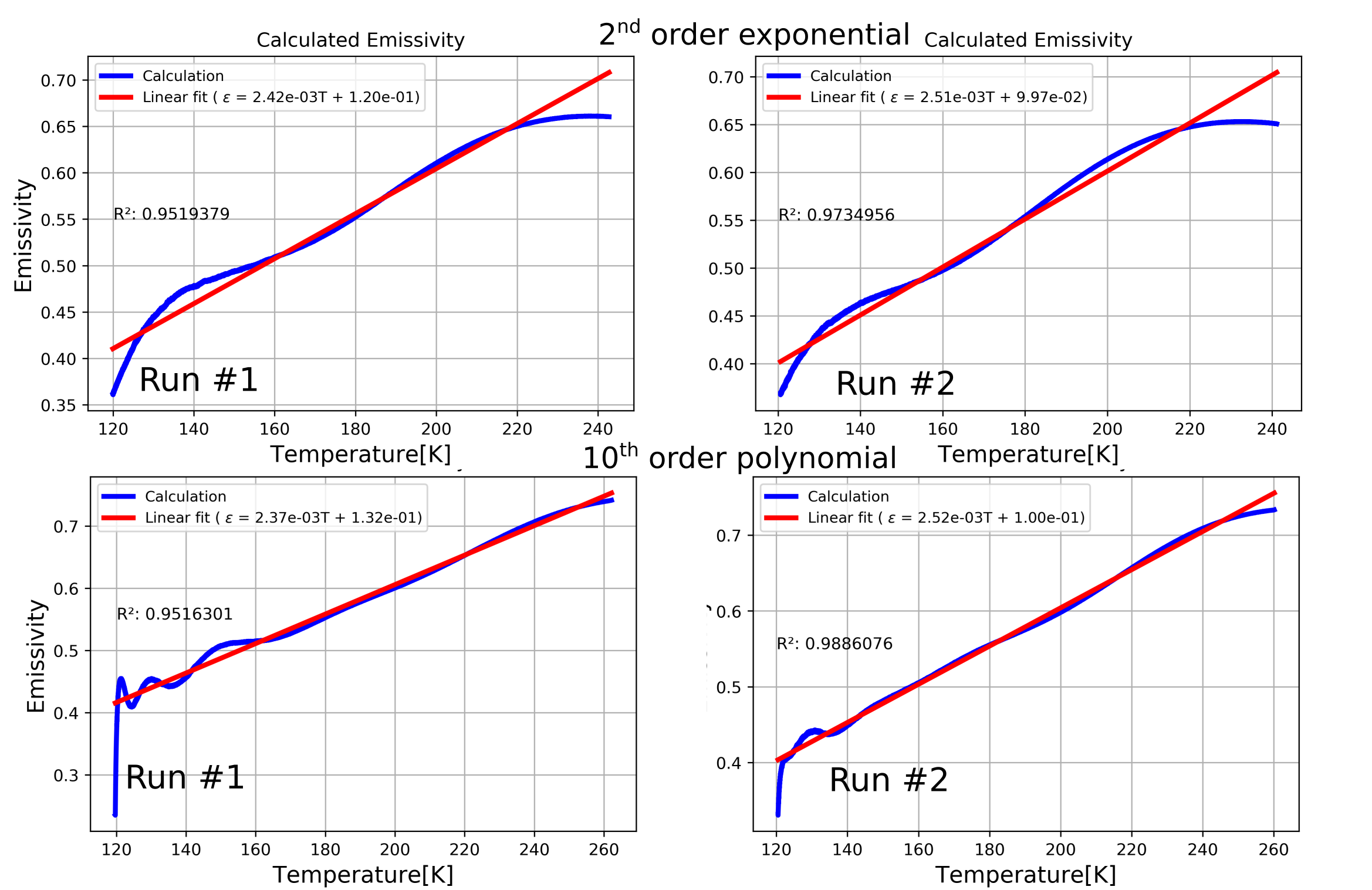}
\caption{Calculated emissivity for runs \#1 and \#2 after curve fitting. On the top, emissivity were calculated from a second order exponential fit and on the bottom they were obtained from a $10^{th}$ order polynomial fit. Polynomial fit is valid from 120-260K.}
\label{fig:e_x_T_alltogether2}
\end{figure}

From the fit, the dT/dt term was easily calculated and then $\epsilon(T)$ was obtained from equation \ref{eq:transient2}.
Figure \ref{fig:e_x_T_alltogether2} shows the $\epsilon(T)$ versus T for both fittings and for both runs. On the top, the result of the second order exponential fit is shown for both runs. On the bottom, the same result is shown for the $10^{th}$ order polynomial.
For the exponential fit, the linear regression between $\epsilon(T)$ and T gives $\epsilon(T) = 2.47\times10^{-3}$T $ + 1.10\times 10^{-1}$, On the other hand, the $10^{th}$ order polynomial fit results in $\epsilon(T) = 2.45\times10^{-3}$T $ + 1.16\times 10^{-1}$.

\subsubsection{Forth approach: numerical differentiation by finite difference approximation}
\label{subsub:ForthApproach}

Finally, the emissivity was also calculated from raw data, without any curve fitting or averaging in the temperature array. However, to overcome fluctuations in the data, the derivative took into account one point in the $(i-n)^{th}$ position before and one point in the $(i+n)^{th}$ position after the evaluated point. It means that for a given temperature $T_i$, the ``derivative'' was calculated as the difference $\frac{\Delta T}{\Delta t} = \frac{T_{i-n} - T_{i+n}}{t_{i+n} - t_{i-n}}$. 

\begin{figure}[!ht]
\centering
\includegraphics[width=1.0\linewidth]{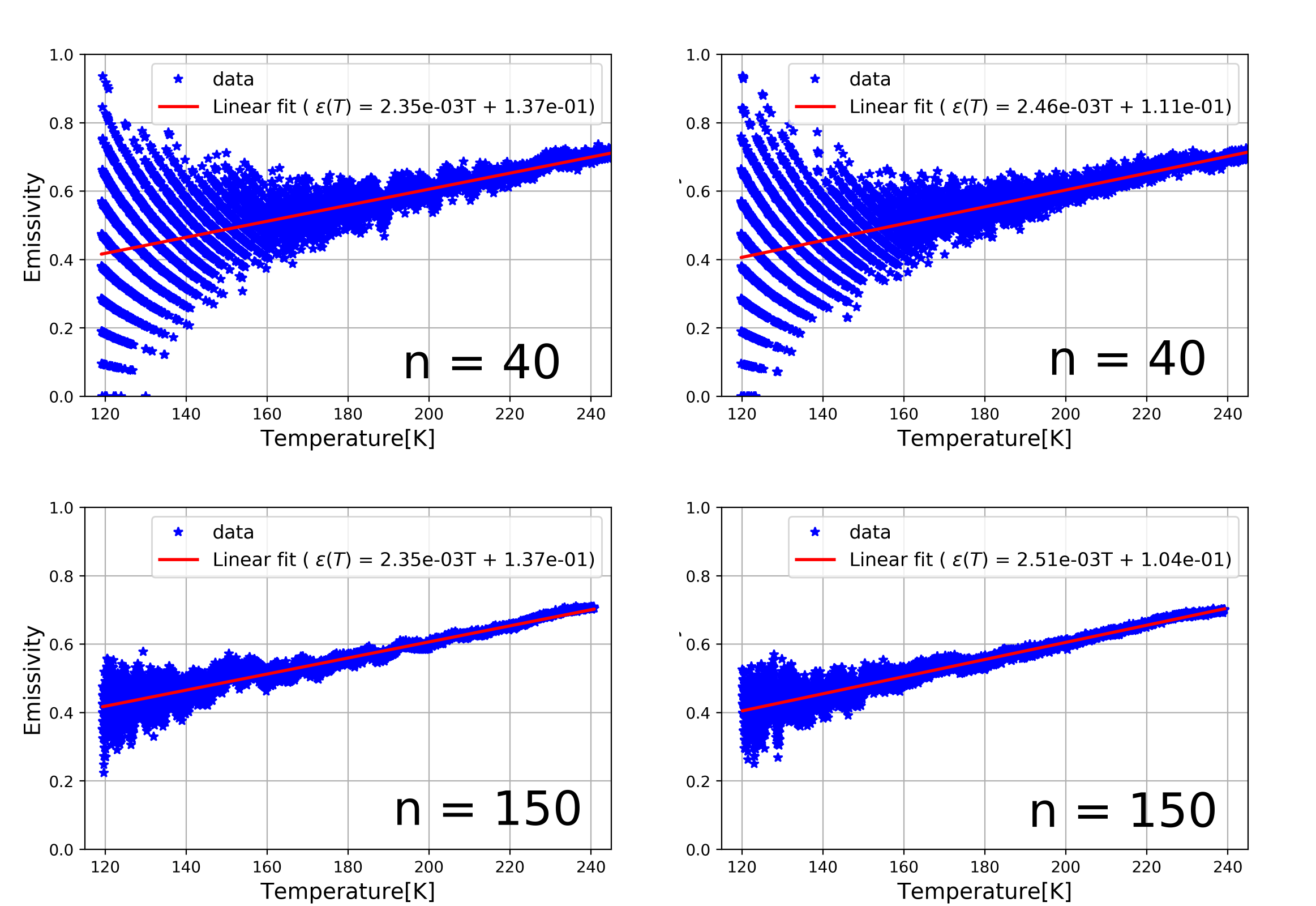}
\caption{Calculated emissivity for run \#1 (left) and \#2 (right) from raw data, without curve fitting nor smoothing (by averaging) in the temperature array. The derivative was calculated by taking into account one point in the $(i-n)^{th}$ position before and one point in the $(i+n)^{th}$ position after the evaluated point.}
\label{fig:emissivity_NO_average_run12_simplified}
\end{figure}

This approach has been applied to a set of different values of \textbf{n} for runs \#1 and \#2. For all cases, the linear regression only had a maximum difference of 2\% and 6.5\% for angular and linear coefficients, respectively. However, for low \textbf{n}, the curve becomes noisier since it is more sensitive to the fluctuations of temperature. Figure \ref{fig:emissivity_NO_average_run12_simplified} shows the graphs for low and high values of \textbf{n} for both runs. A mean value calculated between both runs using $\textbf{n} = 150$ gives $\epsilon(T) = 2.43\times10^{-3}$T $ + 1.21\times 10^{-1}$.

\subsubsection{Summary of results}
\label{subsub:summary}

All the values of emissivities with their respective coefficients of determination ($R^{2}$) for the various data analysis approaches are in Table \ref{tab:dados}.

\begin{table}[ht]
\begin{tabular}{cc|c|c}\hline\hline
\multicolumn{2}{c|}{Data analysis approach}& $\epsilon (T)$ & $R^{2}$\\ \hline
\multirow{2}{*}{\makecell{Savitzky-Golay (SG) \\ filtering}} & Run \#1 & $2.32 \times 10^{-3}$ T $ +~1.42\times 10^{-1}$&0.8832197 \\
                    & Run \#2    &  $2.52 \times 10^{-3}$ T $ +~1.01\times 10^{-1}$  & 0.9065968\\ \hline
\multirow{2}{*}{\makecell{Averaging the digital \\readings}} & Run \#1 & $2.27 \times 10^{-3} $ T + 1.31 $\times 10^{-1}$& 0.9654281 \\
                    & Run \#2    & $2.44 \times 10^{-3} $ T + 9.62 $\times 10^{-2}$    & 0.9608459 \\ \hline
\multirow{2}{*}{\makecell{Curve fitting \\(exp. $2^{nd}$ order)}} & Run \#1 & $2.42 \times 10^{-3} $ T + 1.20 $\times 10^{-1}$ & 0.9519379\\
                    & Run \#2    & $2.51 \times 10^{-3} $ T + 9.97 $\times 10^{-2}$   & 0.9734956\\ \hline
\multirow{2}{*}{\makecell{Curve fitting \\(pol. $10^{th}$ order)}} & Run \#1 & $2.37 \times 10^{-3} $ T + 1.32 $\times 10^{-1}$ & 0.9516301\\
                    & Run \#2    &  $2.52 \times 10^{-3} $ T + 1.00 $\times 10^{-1}$   & 0.9886076\\ \hline
\multirow{2}{*}{\makecell{Finite difference \\ approximation}} & Run \#1 & $2.35 \times 10^{-3} $ T + 1.37 $\times 10^{-1}$  & 0.8590187 \\
                    & Run \#2    & $2.51 \times 10^{-3} $ T + 1.04 $\times 10^{-1}$  & 0.8661256\\
\hline\hline
\end{tabular}
\caption{Summary of results}
\label{tab:dados}
\end{table}

\begin{figure}[!ht]
\centering
\includegraphics[width=0.7\linewidth]{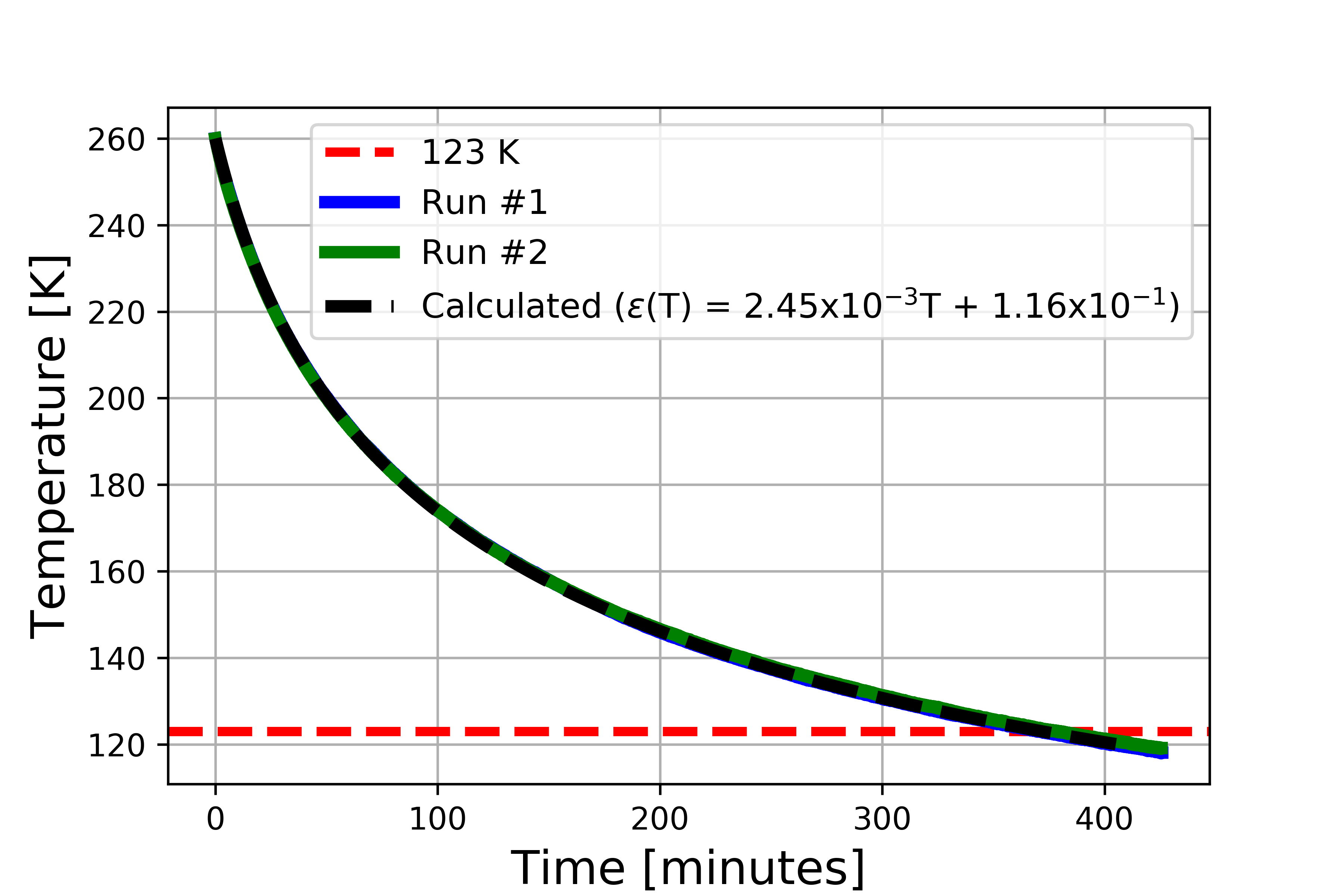}
\caption{Experimental versus the theoretical cooldown curve. Experimental data for runs \#1 (solid blue) and \#2 (solid green) plotted against the theoretical calculation using $\epsilon$(T) = $2.45 \times 10^{-3} T + 1.16 \times 10^{-1}$. This value for $\epsilon$(T) is a mean between the results obtained from the $10^{th}$ order polynomial fit for both runs.}
\label{fig:Exp_calc_polyfit}
\end{figure}

In a general sense, the results obtained through all the methods described here are in good agreement with each other, where the greater difference between the average and the extremes was 6.5\% and 20\% for the angular and linear coefficients, respectively. Although their good agreement, one of them seem to have a best fit, which is the $10^{th}$ order polynomial fit. The emissivity values of this curve fit (the average between the one for run \#1 and the one for run \#2) were used to determine a theoretical cooldown curve, which was compared with the experimental data and shown in figure \ref{fig:Exp_calc_polyfit}. This theoretical cooldown curve was calculated using equation \ref{eq:transient1} and assuming arbitrary small steps of time (2 seconds). The new temperature was calculated from the former one after this time step. This calculated temperature would become the initial temperature for the next calculation step and so on, everything starting from the initial temperature.

Figure \ref{fig:Exp_calc_polyfit} shows the experimental data for both experiments (blue and green curves) and the calculated curve considering radiation and $\epsilon(T) = 2.45 \times 10^{-3} T + 1.16\times 10^{-1}$. As we can see, this calculated function $\epsilon(T)$ greatly agrees with the data.

\subsection{Simulation results}

\begin{figure}[!ht]
\centering
\includegraphics[width=0.7\linewidth]{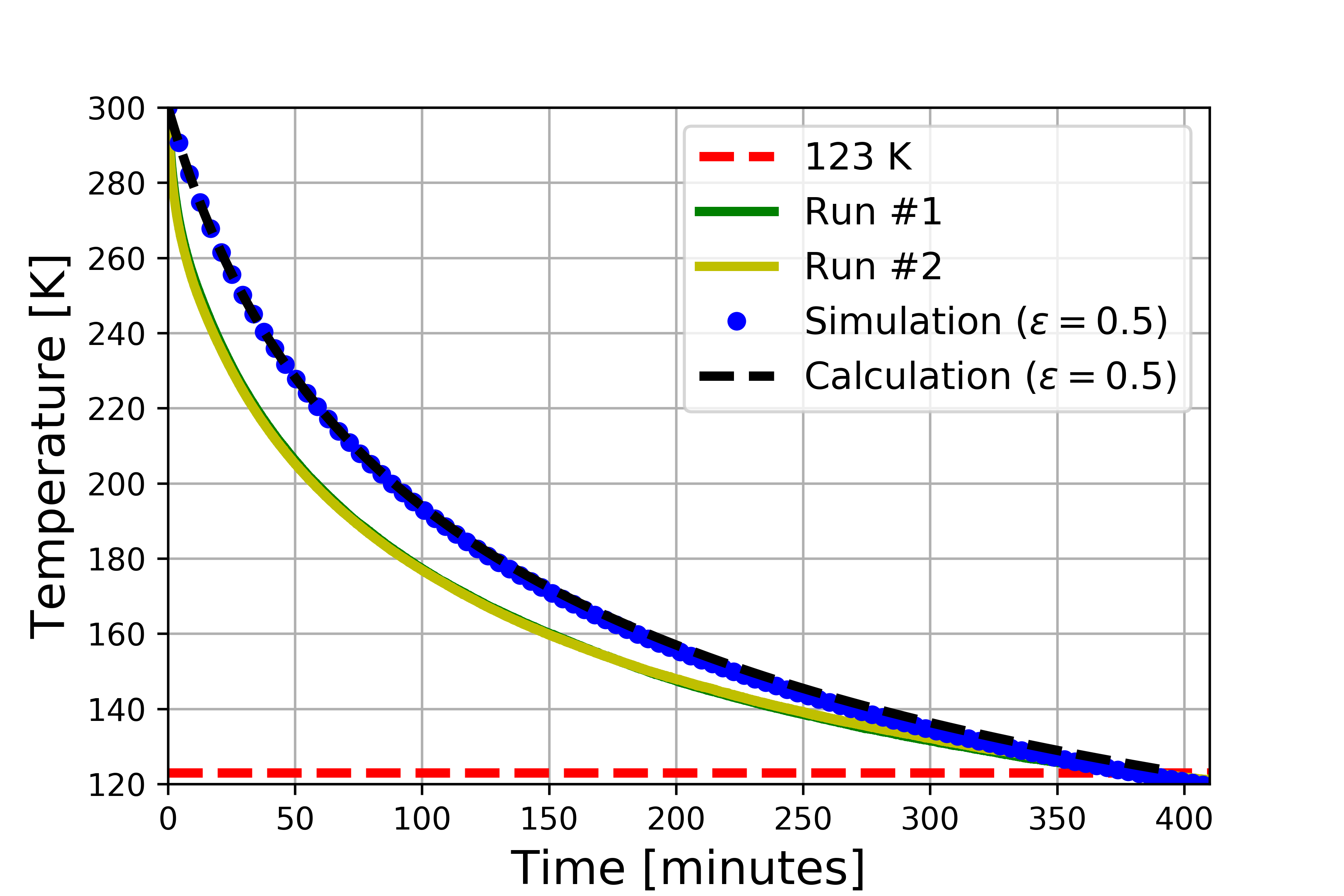}
\caption{Simulation (blue dots) compared to the theoretical calculation (black dashed line), both for a constant emissivity ($\epsilon_{Si}$ = 0.5). Experimental results (run \#1 and \#2, overlapped) are also plotted for comparison.}
\label{fig:Exp_Sim-Calc}
\end{figure}

We performed Solidworks simulation to predict how long the cooling down could last. In a first attempt, a constant emissivity of 0.5 was used. The blue dotted curve in figure \ref{fig:Exp_Sim-Calc} show this temperature drop with time. A calculated curve using only radiation, emissivity of 0.5 and equation \ref{eq:transient1}, satisfactorily predicted the same result, as shown in dashed black curve.

\begin{figure}[!ht]
\centering
\includegraphics[width=0.7\linewidth]{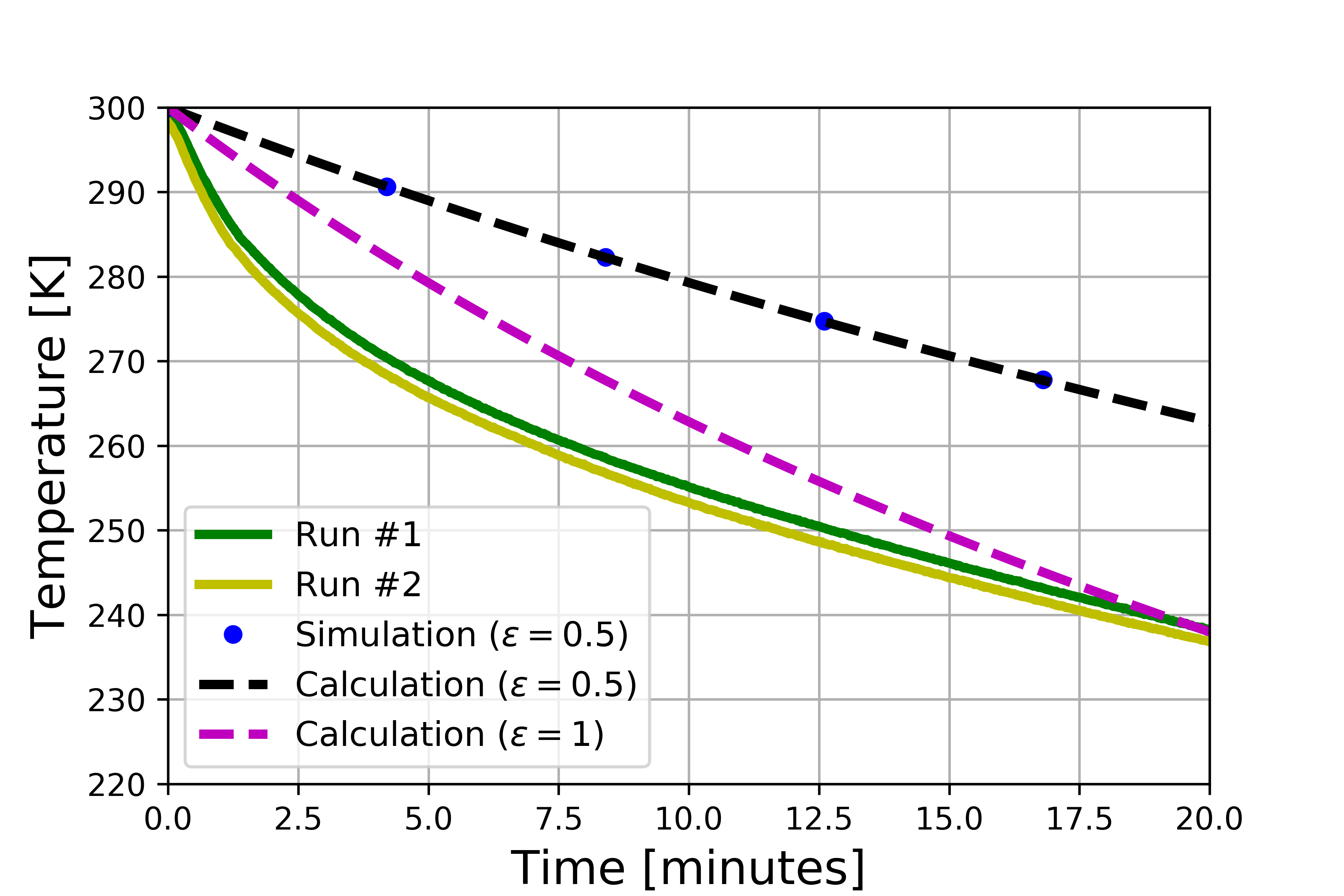}
\caption{Experimental results (run \#1 and \#2, overlapped) plotted against calculation for $\epsilon = 1$ (blackbody) and $\epsilon = 0.5$ and for simulation for $\epsilon = 0.5$.}
\label{fig:Initial_CoolingDownComparison}
\end{figure}

The experimental data for run \#1 and \#2 was also plotted in figure \ref{fig:Exp_Sim-Calc} to make a comparison easier. As can be seen, the cooling down time of the Si sample from room temperature to 123 K was, coincidentally, about the same as it had a constant emissivity of 0.5. Also, the experimental curve for both runs is steeper in the beginning compared to the simulated 0.5 emissivity curve. Actually, in the first 18 minutes, the curve is steeper than the expected for a black body emissivity ($\epsilon = 1$) (as show in figure \ref{fig:Initial_CoolingDownComparison}), which suggests that another heat exchange process can be involved (perhaps due to adiabatic cool-down of the very low pressure air surrounding the Si sample). For this reason, the initial data was not considered for the emissivity estimation and only data below 240 K was used.

\subsection{Comparison to theoretical expectations}
\label{sub:theo_expectation}

\begin{figure}[!ht]
\centering
\includegraphics[width=0.7\linewidth]{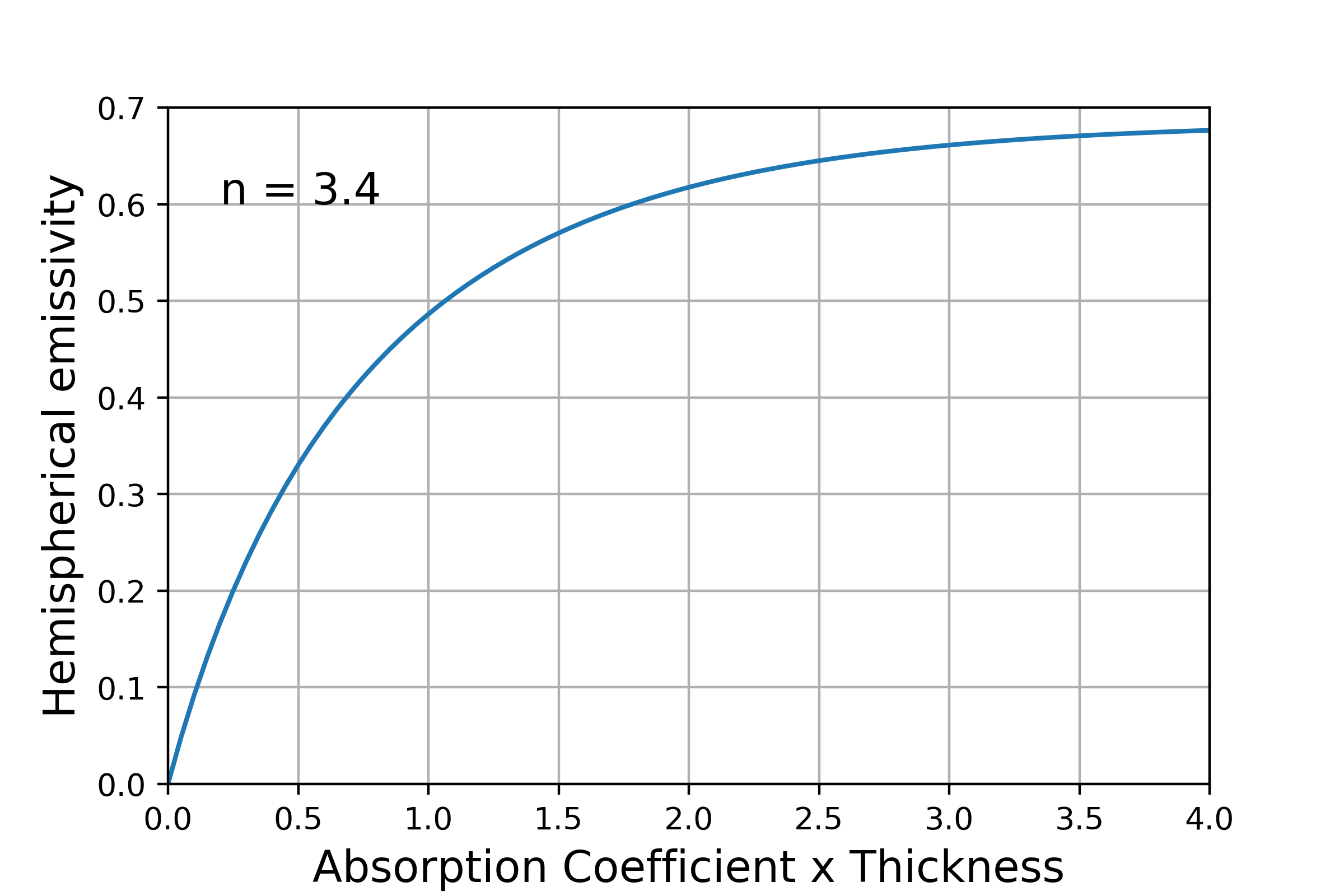}
\caption{Hemispherical emissivity as a function of the product thickness $\times$ absorption for a refractive index n $=$ 3.4 (same as for silicon).}
\label{fig:hem_emis_n3.4}
\end{figure}

Silicon is semi-transparent at wavelengths near to the peak of the black-body spectrum around 123 K. It means that at these wavelengths, the emissivity becomes a bulk instead of a surface phenomenon. Accordingly to Gardon, R.,1956\cite{Gardon1956}, for a given semi-transparent material with refractive index $n$, the hemispherical emissivity is related to the product thickness $\times$ absorption, also termed ``dimensionless thickness''. This relation is shown in figure \ref{fig:hem_emis_n3.4} for $n=3.4$ (same refractive index as silicon).

In order to compare the effective emissivity presented in this paper with a theoretical predicted considering the hemispherical emissivity from a transparent sample, the next subsections are dedicated to the calculation of the absorption and the thickness, and further discussions. 

\subsubsection{Absorption}
\label{subsub:absorption}

As mentioned above, the sample used in this experiment is made of undoped, magnetic Czochralski grown silicon. Free carrier concentration ($n_{i}$) in intrinsic (undoped) silicon is temperature-dependent and has a value of about $(9.7 \pm 0.1) \times 10^{-9} cm^{-1}$ at 300K and $2.3 \times 10^{-6} cm^{-1}$ at 120 K \cite{Misiakos93}. This means that, specially at low temperatures, such low carrier densities makes the free carrier absorption negligible comparable to multiphonon absorption.
The absorption provided by the multiphonon contribution can be expressed as:

\begin{equation}
\alpha = 4 \pi k/\lambda    
\label{eq:alfa}
\end{equation}

where, $\lambda$ is the wavelength and $k$ is the extinction coefficient. Extinction coefficient values for Silicon for different temperatures as function of the wavenumber can be found in \cite{reading_Silicon}. Nevertheless, since we are interested in temperature which is not plotted, we assumed that 123 K would have a mean value between the 100 K and 150 K curves. Also, instead of considering  extinction coefficient only at the peak wavelength (24 $\mu$m, at 123 K), an average value was used, considering a range of wavelengths around the black-body peak, weighted by the black-body curve. This can be summarized in the following equation:

\begin{equation}
    k = \frac{\int_{\lambda_1}^{\lambda_2} k(\lambda) B(\lambda, T) d\lambda }{\int_{\lambda_1}^{\lambda_2} B(\lambda, T) d\lambda}
\end{equation}

where $B(\lambda,T)$ is the black-body curve and k($\lambda$) is the extinction coefficient as a function of the wavelength. For a sample at 123 K, the wavelength peak is about 24 $\mu$m and a range from 16.7 $\mu$m to 33.4 $\mu$m were used to weight the function. Through this approach, we obtained a value of $k = 9.34 \times 10^{-5}$. Finally, an absorption of 0.489 $cm^{-1}$ was estimated from equation \ref{eq:alfa}.

\subsubsection{Thickness}
\label{subsub:thickness}

Since the sample has a rectangular prism shape
with dimensions 7 cm $\times$ 3 cm $\times$ 1.035 cm, we calculated a mean thickness for the whole sample. It was done through the calculation of several paths passing though the sample, in a different set of angles. 
A mesh of 1 mm was created in all surfaces of the sample. The mean thickness was calculated by calculating the path between nodes in different surfaces. For each node in a surface, the path was calculated for all nodes in another surface. Finally, all the paths were summed and divided by the total number of paths. From this approach, we estimated a mean thickness ($\chi$) of 1.67 cm.

\subsubsection{Hemispherical emissivity}
From the calculated absorption and thickness described above we were able to estimate the hemispherical emissivity accordingly to the plot shown in figure \ref{fig:hem_emis_n3.4}. The product $\alpha \chi = 0.489 cm^{-1} \times 1.67 cm  = 0.817$, which corresponds to a hemispherical emissivity of 0.44. At the same temperature, the result calculated in this paper is about 0.41.

Both results have an agreement of about $93\%$, which suggests that the effective emissivity found in this paper could be attributed to the effects of absorption in a transparent media.

\subsection{Discussion and implication of these results}

The results discussed here differs from those used by Weiss\cite{Weiss_tech} in his calculations. Those results were estimated from the silicon infrared absorption weighted by the blackbody spectrum as a function of temperature and are significantly smaller than our results.
Also, since Silicon is semi-transparent at wavelengths near the peak of the black-body spectrum around 123 K, the emissivity will not be an intrinsic property which depends only on the area of the emitter, but a volumetric phenomenon which takes into account the coefficient of absorption and thickness of the sample\cite{Gardon1956} as discussed in subsection \ref{sub:theo_expectation}.
Therefore, further investigations should be conducted in order to confirm this relation between Si emissivity and its sample thickness and/or surface finishing.

Using the same approach used by Rainer Weiss\cite{Weiss_tech}, one can estimate from the results of Uzakbaiuly, B. et al \cite{Uzakbaiuly2018} the emissivity of silicon as a function of temperature and compare them with the present results.

\begin{figure}[h]
\centering
\includegraphics[width=1\linewidth]{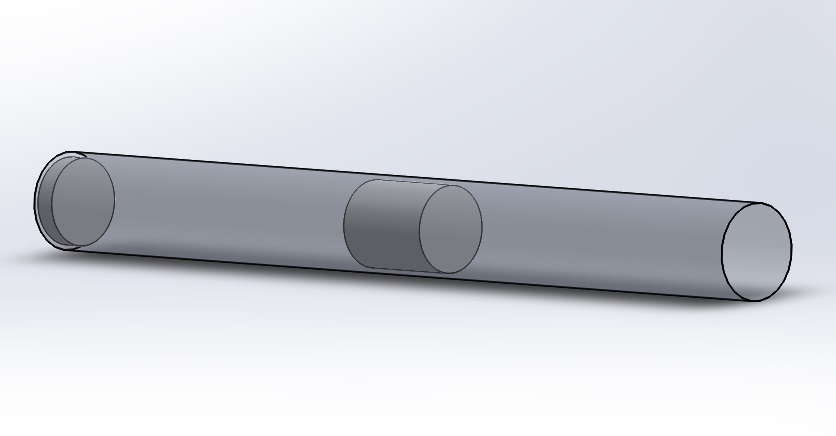}
\caption{ITM sketch. A long tube kept at 60 K shields the ITM. In one end there is a cold compensation plate while the other end is open.}
\label{fig:ITM}
\end{figure}

A direct implication of the results shown in this paper is that, by extrapolating the data up to 300 K, we can predict how long a 200 kg Silicon-made test mass would take to reach 123 K. Voyager's test masses are being planned to be a single cylindrical shaped piece of 450 mm in diameter and 550 mm in thickness/length. For the Input Test Mass (ITM), it has been planned the use of a long shield, maintained at 60 K around it, as shown in figure \ref{fig:ITM}. 

From this configuration and with the Silicon emissivity calculated in this paper we simulated that it's possible to cool the test mass down to 123 K in about 6 days\footnote{Just for comparison, a black coating with $\epsilon = 0.95$ takes about 3 days to do so.}.
This is an important result compared to other options for the initial cool down as described below:

\begin{itemize}
    \item A heat switch could cool the test mass down even faster, however, moving parts touching the test mass can bring risks and implementation challenges.
    \item Exchange gas is another option for the initial cool down, however, after the aimed temperature is achieved, the chamber must be pumped down. For aLIGO, it takes about 2-3 weeks\cite{Voyager_Upgrade} to achieve the target pressure and open the valve to the arms. Assuming that for LIGO Voyager this time would not be substantially reduced, it ends up increasing the amount of time between start of pumpdown and interferometer operation. Also, using gas means that the whole chamber will be cooled, even parts which are not supposed to be exposed to low temperatures. 
\end{itemize}

Although this is a great result for the initial cool down, it is still a concern regarding the steady-state temperature when the laser is locked. This happens because from the calculated curve, we can predict that about 3.5 W can be removed by the barrel at 123 K if the shield is at 80 K and 4.0 W if the shield at 60 K. If we consider that both faces have an emissivity of 0.5, this adds up to 1.9 W. This means that around 5 to 6 W of power can be removed by the test mass (barrel + faces) depending on the temperature of the shield. This restricts the amount of heat which can be absorbed by the coating and by the substrate.
On the other hand, if the theory of semi-transparent Silicon shows itself to be valid, it means that the LIGO Voyager test mass could show an emissivity of up to 0.7 at 123 K, which means that the total amount of heat that could be extracted by the test mass (considering that both faces have $\epsilon = 0.5$) is about 8.6 W(7.5 W) for a  shield at 60 K(80 K). Just for comparison, a black coating could increase the barrel emissivity to about 0.95 and would lead to a heat removal of around 11.0 W(9.6 W) for a shield at 60 K(80 K). This is an improvement of about 22\%. It is important to keep in mind that avoiding the use of any additional coating will minimize Brownian noise.

See \ref{sec:SteadyState_esti} for details.

The discussion presented here shows how important is the sample thickness when the emitting medium is transparent. A few Silicon samples are being prepared at the time this paper is being written and the authors hope to be able to perform these measurements as early as possible.

\section{Conclusions}
\label{Sec:conclude}

In this paper we presented the temperature-dependent effective emissivity of a silicon sample, estimated from its cool-down curve in a constant low temperature environment (~82K). The emissivity value follow the linear dependency $\epsilon(T) = 2.45 \times 10^{-3} T + 1.16 \times 10^{-1}$ in the 120-260 K temperature range. This result is of great interest to the LIGO Voyager gravitational wave interferometer project since it would mean that no extra high thermal emissivity coating on the test masses would be required in order to cool them down to 123 K. The results presented here indicate that bulk silicon itself can have sufficient thermal emissivity in order to cool the 200 kg LIGO Voyager test masses only by radiation in a reasonable short amount of time (less than a week). However, it is still not clear if the natural emissivity of silicon will be sufficient to maintain the LIGO Voyager test masses at the desired temperature (123 K) while removing power absorbed by the test masses. With the present results, a black coating on the barrel surface of the test masses would be necessary if power in excess of 6W is delivered. However, the agreement we found between the hemispherical emissivity obtained by a theory of semi-transparent Silicon and the result obtained in this paper makes us believe that the LIGO Voyager test masses, because of their dimensions, will have effective emissivities around 0.7, which would be enough to remove about 8.6 W (7.5 W) for a shield at 60 K (80K), and the cool-down time would be even shorter than 6 days. This hypothesis may be confirmed in the near future with new measurements.

In any case, further investigations need to be done in order to evaluate the total amount of heat that will be deposited on the LIGO Voyager test masses. The initial runs of LIGO Voyager could be done with no black coating on the barrel parts and check if this would be sufficient for the maximum power achieved in the laser cavities during the run. If necessary, a black coating would be applied for the subsequent runs.
 
\section{Acknoledgements}
\label{sec:acknoledgment}

MC and ODA thank CNPq for financial support (grants \#300240/2019-8 and \#302841/2017-2). Also, they thank to Manel Molina Ruiz for the useful discussion about fitting and to Martin Fejer for suggesting a way to compare the experimental results to theoretical expectations.

 \appendix
\section{Steady-state estimation}
\label{sec:SteadyState_esti}

In this section we present the steady-state calculation for a test mass of 450 mm in diameter and 550 mm in thickness (h). The $A_{Barrel}$ is $2 \pi r h$ and the area of the face is given by $\pi r^{2}$. An emissivity of 0.5 was used for the faces. We performed the calculation for three different barrel emissivity values:

\begin{enumerate}
    \item $\epsilon = 0.42$, from the expression found in this paper ($\epsilon(T) = 2.45\times10^{-3}T + 1.16\times10^{-1}$), for $T = 123$ K.
    \item $\epsilon = 0.70$, from the semi-transparency effect as described in the paper;
    \item $\epsilon = 0.95$, from a theoretical black coating.
\end{enumerate}

Since the face has a constant emissivity ($\epsilon = 0.5$), the amount of irradiated heat is calculated as
\begin{equation}
P_{face} = 2\times[\epsilon_{face}A_{face}\sigma(123^4 - 80^4)] = 0.5(\pi(0.225)^2)\sigma(123^4 - 80^4)] = 1.7~W    
\end{equation}

for a shield at 80 K and,
\begin{equation}
P_{face} = 2\times[\epsilon_{face}A_{face}\sigma(123^4 - 60^4)] =  2\times[0.5(\pi(0.225)^2)\sigma(123^4 - 60^4)] = 1.9~W    
\end{equation}

for a shield at 60 K.

\subsection{\textbf{$\epsilon = 0.42$}}
\label{subsec:case1}

For $\epsilon = 0.42$, the barrel is able to remove:
\begin{equation}
P_{barrel} = \epsilon(123 K)A_{barrel}\sigma(123^4 - 80^4) =
0.42(2\pi(0.225)(0.55))\sigma(123^4 - 80^4) = 3.5~W
\end{equation}

for a shield at 80 K and,
\begin{equation}
P_{barrel} = \epsilon(123 K)A_{barrel}\sigma(123^4 - 60^4) = 0.42(2\pi(0.225)(0.55))\sigma(123^4 - 60^4) = 4.0~W
\end{equation}

for a shield at 60 K. 

Summing the amount of heat irradiated by the barrel and the face we have 5.2 W(5.9 W) for a shield at 80 K(60 K).

\subsection{$\epsilon = 0.70$}
\label{subsec:case2}

For $\epsilon = 0.7$, the barrel is able to remove:
\begin{equation}
P_{barrel} = 0.70A_{barrel}\sigma(123^4 - 80^4) =
0.70(2\pi(0.225)(0.55))\sigma(123^4 - 80^4) = 5.8~W
\end{equation}

for a shield at 80 K and,
\begin{equation}
P_{barrel} = 0.70A_{barrel}\sigma(123^4 - 60^4) = 0.70(2\pi(0.225)(0.55))\sigma(123^4 - 60^4) = 6.7~W
\end{equation}

for a shield at 60 K. 

Summing the amount of heat irradiated by the barrel and the face we have 7.5 W(8.6 W) for a shield at 80 K(60 K).

\subsection{$\epsilon = 0.95$}
\label{subsec:case3}

Finally, for $\epsilon = 0.95$, the barrel is able to remove:
\begin{equation}
P_{barrel} = 0.95A_{barrel}\sigma(123^4 - 80^4) =
0.95(2\pi(0.225)(0.55))\sigma(123^4 - 80^4) = 7.9~W
\end{equation}

for a shield at 80 K and,
\begin{equation}
P_{barrel} = 0.95A_{barrel}\sigma(123^4 - 60^4) = 0.95(2\pi(0.225)(0.55))\sigma(123^4 - 60^4) = 9.0~W
\end{equation}

for a shield at 60 K. 

Summing the amount of heat irradiated by the barrel and the face we have 9.6 W(11 W) for a shield at 80 K(60 K).

\bibliographystyle{model1-num-names}
\bibliography{ms}

\end{document}